\documentclass[aps,twocolumn,prl]{revtex4-1}

\usepackage{natbib}
\usepackage{color}
\usepackage{todonotes}

\usepackage{graphicx}
\usepackage[reqno]{amsmath}
\usepackage{latexsym}
\begin{document}
\title{Robust and adjustable C-shaped vortex beams} 
\author{M. Mousley}
\author{G. Thirunavukkarasu}
\author{M. Babiker}
\author{J. Yuan}
\affiliation{Department of Physics, University of York,Heslington, York YO10 5DD}

\begin{abstract} 
Wavefront engineering is an important quantum technology.  Here, we demonstrate the design and production of a robust C-shaped and orbital angular momentum (OAM) carrying beam in which the doughnut shaped structure contains an adjustable gap.  We find that the presence of the vortex line in the core of the beam is crucial for the robustness of the C-shape against beam propagation.  The topological charge of the vortex core controls mainly the size of the C, while its opening angle is controlled by the presence of vortex-anti-vortex loops.  We demonstrate the generation and characterisation of C-shaped electron vortex beams, although the result is equally applicable to other quantum waves.  Applications of C-shaped vortex beams include lithography, dynamical atom sorting and atomtronics. 
\end{abstract}
\maketitle
Structured quantum waves with a patterned phase or intensity distribution are of theoretical and practical interest in emerging quantum technologies.  To date most of the research interest has focused on optical and electron vortex beams which are characterised by a phase singularity on the beam axis and a cylindrically symmetric intensity distribution, often called a doughnut mode.  The phase singularity is known to be associated with the property of orbital angular momentum (OAM) carried by the beam\cite{Allen1992, Bliokh2007}, which has given rise to prominent applications including nano-manipulation of particles \cite{He1995}, OAM entanglement \cite{Mair2001}, and multiplexed data transfer \cite{Wang2012}.  Research in vortex beams has also stimulated interest in other types of structured quantum waves such as Airy beams which have an interesting 'self-accelerating' property \cite{Voloch-Bloch2013a}. It is clear that there is opportunity to explore other shaped beams, the preparation and characterisation of which as well as their interactions with matter would greatly enrich our understanding of the nature of quantum physics with potential for practical applications.

Here we focus on the controlled formation of C-shaped beams which has many characteristics and useful applications distinct from the beams with a circularly symmetric intensity illumination\cite{Verbeeck2012, McMorran2011}.  For example, orbital angular momentum in a beam devoid of cylindrical symmetry is itself of fundamental interest.  In addition, such a beam can be used in lithography to produce shaped nanostructures, such as those used in the split ring structure of metamaterials research \cite{Marques2002}, without scanning a beam. C-shaped beams are also used in atomtronic quantum interference devices \cite{Eckel2014} and, for the electron version, in the sensing of the Gouy phase and applied magnetic fields \cite{Arlt2003, Hamazaki2006, Cui2012, Guzzinati2013}.

We begin by introducing an analytically precisely defined and robust C-shape vortex beam with an opening gap that can be easily tuned from zero.  We attribute this to the presence of vortex line threading its core, which also enhances the robustness of the C-shaped beam against propagation in free space.  We show here that it carries a well defined OAM despite not having a circularly symmetric cross section.  Finally, we discuss the application of C-shaped electron vortex beams in the fabrication of nanostructures and in the manipulation of nanoparticles or cold atoms for use in atomtronics.  

\begin{figure}[h!]
{\includegraphics[width=\columnwidth]{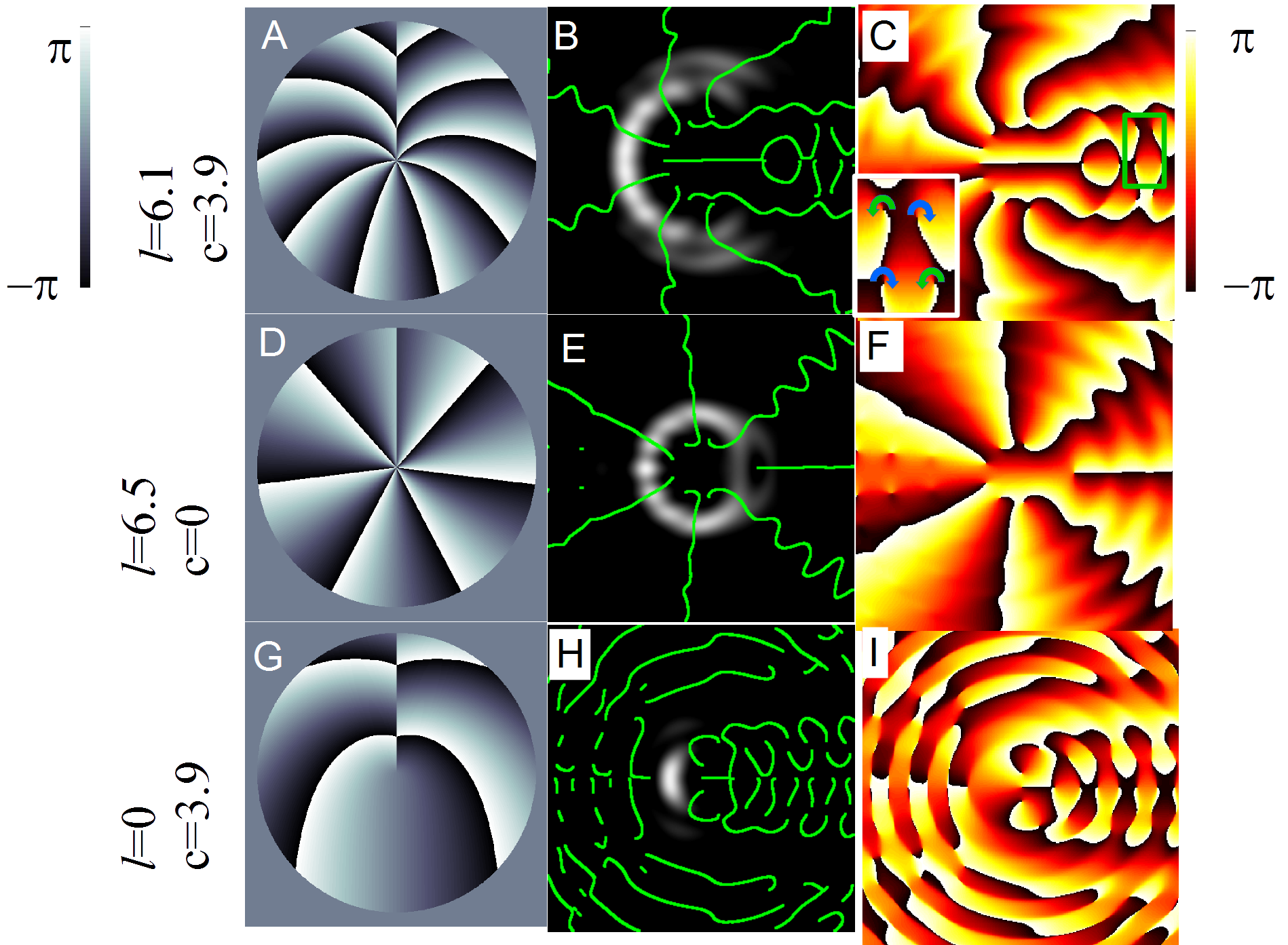}}
\caption{{ (A, D, G): Phase distributions wrapped in unit of $2\pi$ (black = $-\pi$ and white = $\pi$).
(B,E,H): Fourier transform intensity distributions (white is high intensity), with constant phase contours shown in green. (C,F,I): Phase of the Fourier transform  (black=$-\pi$ white=$\pi$ ). The inset in the bottom left of C shows vortices and anti-vortices (blue and green arrows) found in the green rectangle in C. The $l$ and c values for each row are shown on the left hand side.
}}\label{2Dtheory}
\end{figure}

The phase function needed to modulate the wavefunction of an incoming plane waves such that a C-shaped vortex beam can be produced at the far field diffraction plane has a very simple form:
\begin{equation}
(l+c\bar{\rho})\phi \label{mainphase}
\end{equation}
where $l$ and $c$ are adjustable parameters, $\phi$ (= $\tan^{-1}{(-x/y))}$ is the azimuthal angular variable in cylindrical polar coordinates and $\bar{\rho}=\rho/\rho_{max}$ where $\rho$ is the radial variable and $\rho_{max}$ is the limiting radius of a circular aperture function.  An example with $l=6.1$ and $c=3.9$ is shown in Fig. \ref{2Dtheory}.  The phase is evaluated within the range $-\pi$ to $\pi$, so that the phase discontinuity line coincides with the positive $y$-axis, forming a gap along the positive $x$-axis in the far field diffraction plane (Fig. \ref{2Dtheory}B).

The phase function can be understood by examining the two contributions in Eq.(\ref{mainphase}) separately.  The first contribution only has an azimuthal dependence and it is instructive to consider the physical contents of this term first.  With the parameter $l$ an integer, this term gives rise to standard vortex beams whose diffraction patterns consist of a doughnut ring with a topological vortex charge, or winding number,  $l$ at its core.  When $l$ is a fractional number, the phase difference at the start and the end of a $2\pi$ rotation is no longer an integer multiple of $2\pi$, resulting in a sharp azimuthal discontinuity in the phase function.  The far field diffraction of this phase discontinuity creates a localised defect in the doughnut structure of the the vortex beam.  Figures \ref{2Dtheory}D-\ref{2Dtheory}F show, in the way of an example, the case for $l$=6.5 ($c$=0).  These results can be understood within a geometric ray model by considering the local phase gradients as wavevectors of the rays contributing to the intensity of the diffraction pattern. In the case of an integer $l$ vortex beam the uniform phase gradient gives rises to rays with a transverse component uniformly distributed in the azimuthal direction.  An azimuthal phase discontinuity corresponds to the missing rays within certain azimuthal transverse directions.  A C-approximate can be realised when fractional $l$ is half way between the integers \cite{Abramochkin1991a, Berry2004, Leach2004, Gotte2008, Garcia-Gracia2009}. However, due to the interference effect, even in that case there is still significant intensity in the opening, making the fractional vortex beam often not an ideal C-shaped beam (for example, see Fig. \ref{2Dtheory}E).

Here we demonstrate that the deficiency of the fractional OAM state can be minimised by including a spiral phase function term that depends on both the radial and azimuthal coordinates.   Such a phase function, when defined over the interval between 0 and 2$\pi$, has previously been shown to produce a spiralling intensity pattern \cite{Alonzo2005, Hermosa2007, Li2010}.  We have defined the phase function over the interval between -$\pi$ to $\pi$ to produce a C-like symmetric opening in the far field diffraction pattern.  The size of the phase jump at the azimuthal phase discontinuity is now a linear function of the radial coordinate (see Fig. \ref{2Dtheory}). This is a stable feature irrespective of the value of the parameter $l$, allowing the latter to be adjusted freely in defining the size and opening angle of the C -shape as shown below.  The opening due to the phase discontinuity is now much more clearly defined when compared with the results produced by the fractional vortex beam.  Fig. This is associated with a high density packing of local vortex and anti-vortex components in the phase pattern of the far field diffraction as in Fig. \ref{2Dtheory}I.  As the vortices occupy regions of darkness, high density packing is essential for an extended dark region, an essential requirement for a C-shaped beam with a large opening.

The tunability of the C-shaped beam is realised by varying the ratio $l:c$ thus changing the relative importance of the term involving fractional topological charge \cite{Berry2004, Leach2004, Gotte2007} with respect to the term involving radial phase gradients (as seen in optical twists \cite{Daria2011}).  An analysis of the computed phase distributions shows that the resultant beam still retains an overall topological charge at the centre, but now the vortex possesses other topological charges of both signs distributed over a finite area along the line bisecting the gap region.  The density of vortex and anti-vortex pairs is higher than in the case of fractional vortex beam and this density is adjustable.   As the value of $c$ increases the phase distribution at the focal plane shows a gradual stacking of vortex-antivortex pairs into a grid like collection (Fig.\ref{2Dtheory}), producing an opening of increasing size. This occurs through the displacement of the isophase lines for larger $c$ as in Fig.\ref{2Dtheory}F where similar phase lines are seen to occur at larger angles to the negative $x$ axis (upper isophase lines moving 'clockwise' and lower isophase lines moving  'anticlockwise')
( see supplementary material for animation). 

\begin{figure}[h!]
\includegraphics[width=\columnwidth]{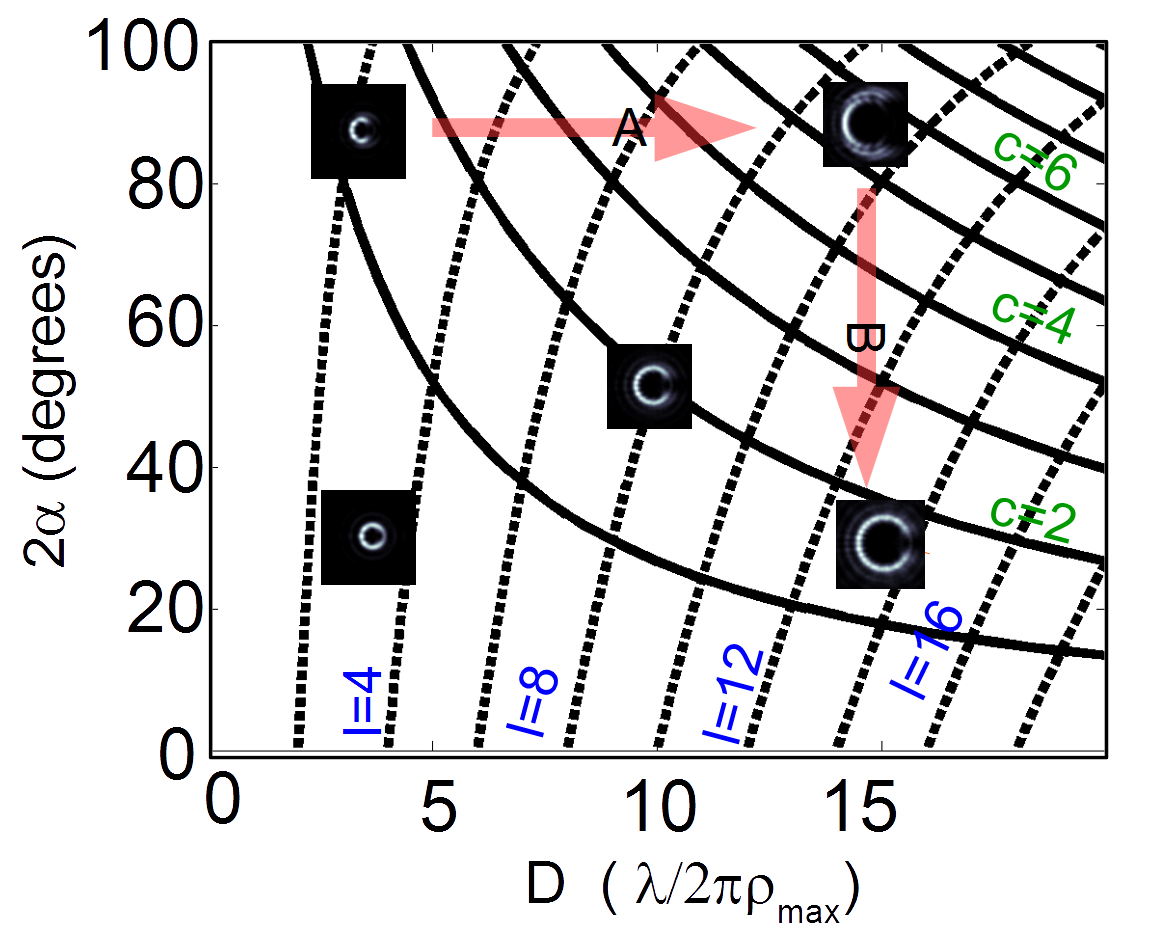}
\caption{A contour plot of parameter values $l$ and c for desired opening angle ($\alpha$ in degrees) and size (D measured in the dimensionless units of $\lambda/2\pi\rho_{max}$, where $\lambda$ is the wavelength). Five examples are shown as inserts according to their positions in the $(D,2\alpha)$ parameter space.  The arrow A shows positions in ($D$,$2\alpha$) plane which produce C-shapes of fixed opening angle but varying sizes conversely the arrow B shows parameters producing a C-shapes of a fixed size with varying opening angles.}
\label{control}
\end{figure}

By controlling both parameters, namely $l$ and $c$, one can alter independently both the size, D (the intersection of the C shape with the negative x axis in reciprocal space), and the opening angle, $2\alpha$, of the C-shape (measured from the reciprocal space origin to the ends of the C) (Fig.\ref{control}).  This versatility leads to the possibility of realising various C-shaped beams such as those with variable size or variable opening angle, as shown by the arrowed paths in the parameter space in Fig.\ref{control} and demonstrated in supplementary material.

\begin{figure}[h!]
\includegraphics[width=\columnwidth]{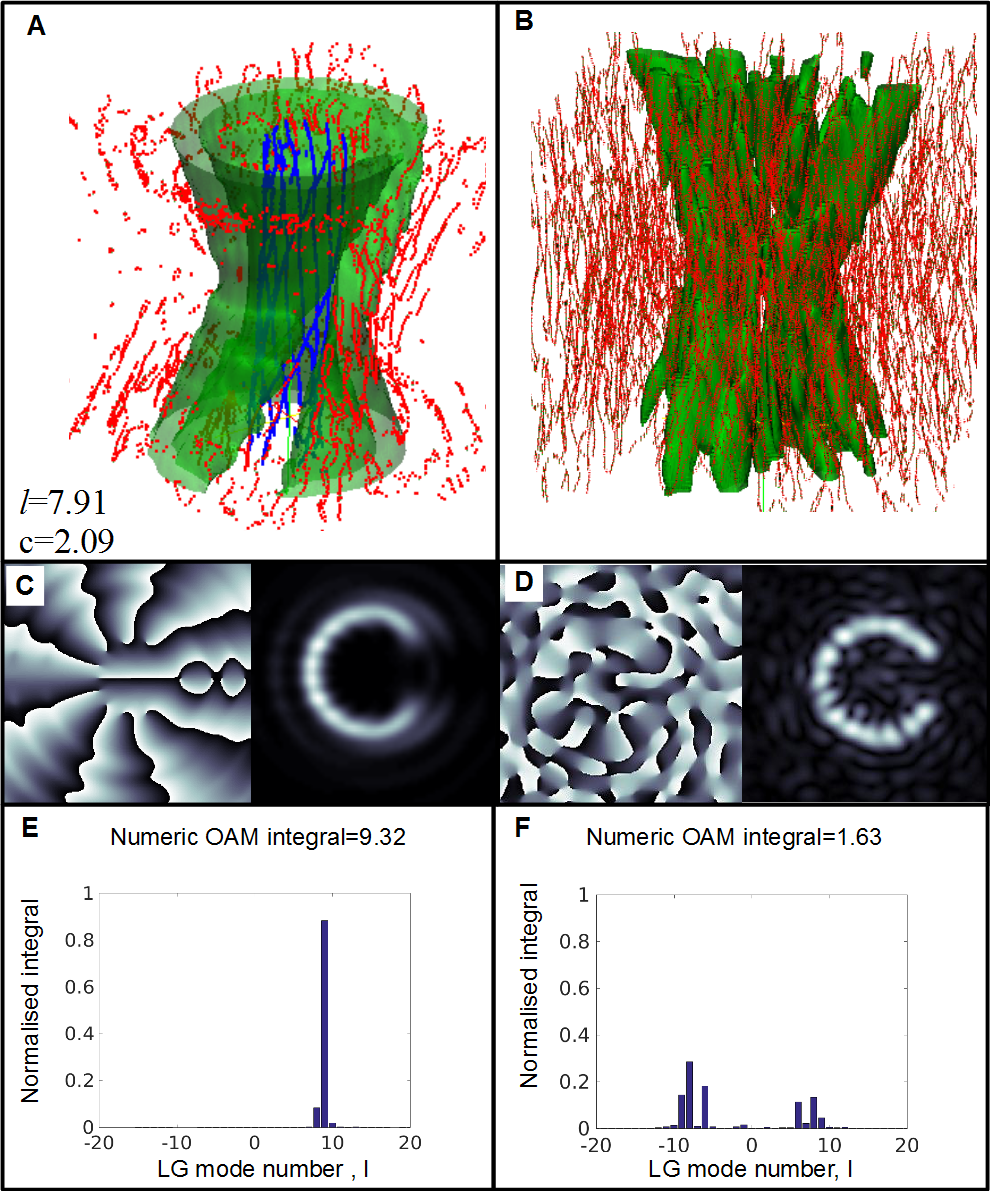}
\caption{ A) A simulated electron C shaped beam using the analytical phase function with $l=7.91$ and $c=2.09$.  The blue lines show the central collection of nodal lines responsible for the OAM and robustness of the C shaped beam.  Other vortex nodal lines are shown in red and the high intensity volume shown in green.  B) As for A but using the phase function from the IFTA result. The simulated propagation is between $-13.6 \mu m$ to $+13.6\mu m$ about the focus along the z axis, for an electron beam with $\lambda=2.5\times 10^{-12}$m in an optical system with a camera length of $0.016m$.  C) and D) show the at focus phase (left) and intensity (right) plots for the C-shaped beam shown in C) and D) respectively. E) and F) are the corresponding OAM mode decompositions obtained by performing numerical overlap integrals using the Laguerre-Gaussian (LG) basis set of modes of varying azimuthal quantum number for  a radial quantum number of 1.}  
\label{3Dtheory}
\end{figure}

Our approach to beam shaping can be compared with the alternative approach of phase mask production using an Iterative Fourier Transform Algorithm (IFTA)\cite{Shiloh2014a}. We have carried out an IFTA calculation involving 2000 iterations with a C-shaped target intensity distribution and the results are shown in Fig, \ref{3Dtheory}. As can be seen, both methods produce C shaped intensities, however the IFTA intensity result had undesirable fine structure visible in the intensity distribution around the C shape.  This is a known issue with the IFTA approach \cite{Shiloh2014a} and is due to the pixelated nature of the derived phase mask, with its characteristic randomly distributed phase singularities.

An examination of the propagation of the phase structure shows that there are not only far more numerous vortex lines (as shown by the red lines in Fig.\ref{3Dtheory}B) but also that many of these vortex lines are dislocation loops, a feature characteristic of the speckle pattern seen in Fig.\ref{3Dtheory}D.  In contrast to the IFTA-derived phase function, our analytical phase functions are smooth over the aperture plane except near the phase discontinuity.  This focuses the number of vortex-anti-vortex dislocation loops only in the regions where they are required to generate the opening.  As a result, our C shaped beam has a very simple collection of phase singularities which propagate along with the beam inside the high C-shaped intensity pattern  (blue lines in figure \ref{3Dtheory} A). Furthermore, the nodal lines are seen to be slightly displaced from a perfect circle, but they still propagate along the beam axis.  The combined effect of this and nodal loops at larger radii, creates the opening seen in the green volume of Fig.\ref{3Dtheory}. Thus the C-shaped structure is robust against beam propagation apart from a slow rotation which can be attributed to the Gouy phase change at the beam waist \cite{Guzzinati2013}. 
  
A simple and direct manifestation of the very different vortex structures of the two types of C-shaped vortex beams can be seen in their OAM characteristics.  With its simple vortex structure at the beam axis, our C-shaped beam has a chiral phase structure and a narrow OAM mode distribution centred around $l=9$ as shown in Fig.\ref{3Dtheory}E.  The net OAM content can be calculated analytically from the phase function given in Eq.(\ref{mainphase}) by $\frac{\int{\psi^* \hat{L}_z\psi}}{\int{\psi^*\psi}}=(l+{[2c/3]})\hbar\label{expec}=9.3\hbar$ with $l=7.91$ and $c=2.09$.  This equals to the OAM numerically calculated for the vortex beam at far field, indicating the conservation of OAM in a lens system.  The contribution from the spiral phase term is due to the shift of the barycenter of the beam off the axis.  The C-shaped intensity produced by the IFTA method, however, has a much smaller net OAM of $1.6\hbar$.  This shows that the C shaped vortex beams has clear advantages in applications where OAM is desired, such as in particle manipulation experiments. In pure vortex beams, OAM of the beam is associated with the circular circulation and is connected with the cylindrical symmetry of the transverse structure.  The existence of net OAM content in a non-symmetric intensity distribution is nontrivial but can only be understood as a quantum mechanical effect due to the coherent superposition of several OAM states such that the local intensity minimum can be the result of destructive interference between the OAM modes.

Our results are applicable to all beams that can be described by scalar quantum waves.  In the following, we first present the experimental demonstration of the C-shaped electron beam formation inside an electron microscope, although the result should be equally valid for the shaping of wavefronts of other quantum waves such as electromagnetic and other matter waves.  

\begin{figure}[h!]
\includegraphics[width=\columnwidth]
{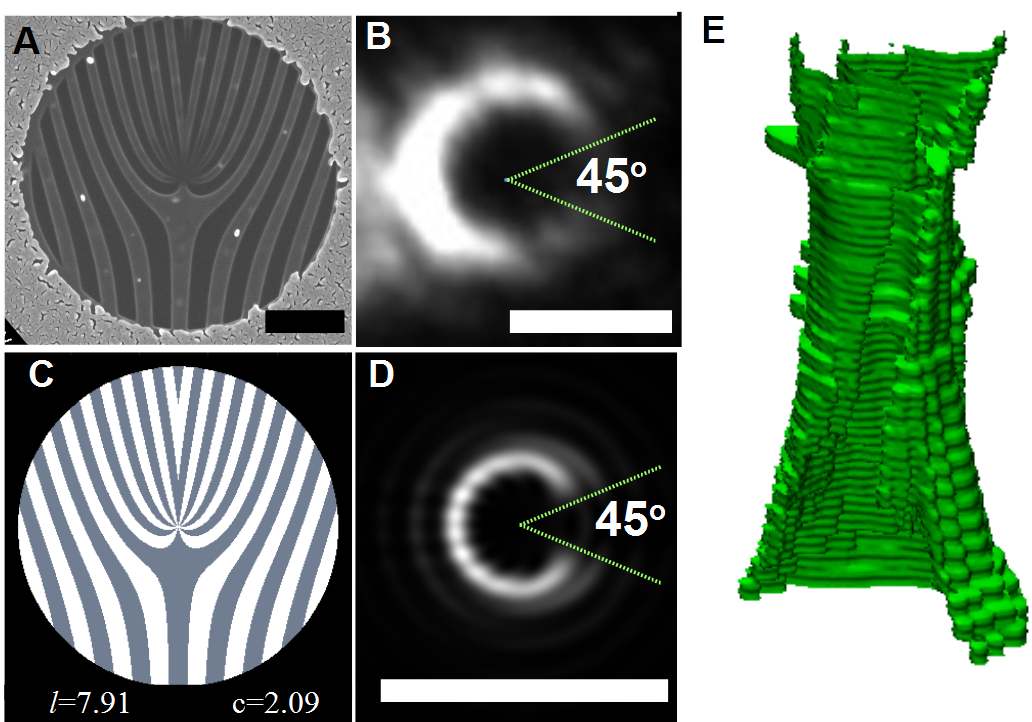}
\caption{ A) a scanning electron microscopy image of the produced mask; B) focal plane experimental intensity distribution, with the green lines highlighting the $45^o$ angle; C) the calculated hologram design used to create the mask; D) the simulated intensity distribution of the target C shape; E) experimental intensity distributions at different $z$-planes, with the green volume being the volume of high intensity.  (scale bars:A)1um,B)500um on CCD, D)1urad)}
\label{exp}
\end{figure}

We chose to encode our desired phase function in a binary computer generated hologram pattern (see Fig.\ref{exp}C, for an opening angle $2\alpha= 45$, $D= 10 \lambda/2\pi\rho_{max}$, by setting $l=7.91,c=2.09$.  The binary hologram pattern is then transferred as a thickness pattern using focused ion beam (FIB) milling, onto a silicon nitride membrane of $200$ nm with a $50$ nm Pt/Pd layer. The Pt/pd layer is removed to form a circular aperture and the mask pattern is then milled to create thickness variation inside the aperture. As the total sample thickness of the mask is $\le 200nm$ , it is possible to neglect absorption as a first approximation and the resulting structure, when illuminated by a coherent electron wave, acts as a binary phase mask due to interactions with the mean inner potential of the film \cite{Shiloh2014a, Grillo2014a}. 
 
Figure \ref{exp}B displays the far field Fraunhofer diffraction pattern of the mask in Fig. \ref{exp}A produced in an electron microscope, operated in free lens control and with operating voltage of 200 $kV$ ( $\lambda=2.5\times 10^{-12}$m). The result shows a strong match with the corresponding simulation shown in Fig.\ref{exp}D, demonstrating a successful control of the C structured illumination including the opening angle $2\alpha=45^o$. We have found that the intensity of the lower section of the illumination from the phase mask is less than expected.  This is most likely due to imperfections in the mask caused by the inherent stochastic nature of the milling process.

The robustness of the C-shaped intensity against beam propagation is evident in the preservation of the pattern as a function of defocus (Fig.(\ref{exp})E).  This is similar to the result of simulation shown in Fig.\ref{3Dtheory}, apart from an artefact due to overlapping with the edge of the zero-order beam at large defocus.  

Such a C-shaped electron vortex beam can be directly used in coherent electron beam lithography, for example, producing nanostructures by exposing direct writing electron beam resists to such beams \cite{Mousley2015}.  The advantage is that no scanning beam is used.  We are also exploring the possibility to measure magnetic field by field induced rotation of the C-shape vortex beam \cite{Guzzinati2013, Greenshields2012}.

\begin{figure}[h!]
\includegraphics[width=\columnwidth]{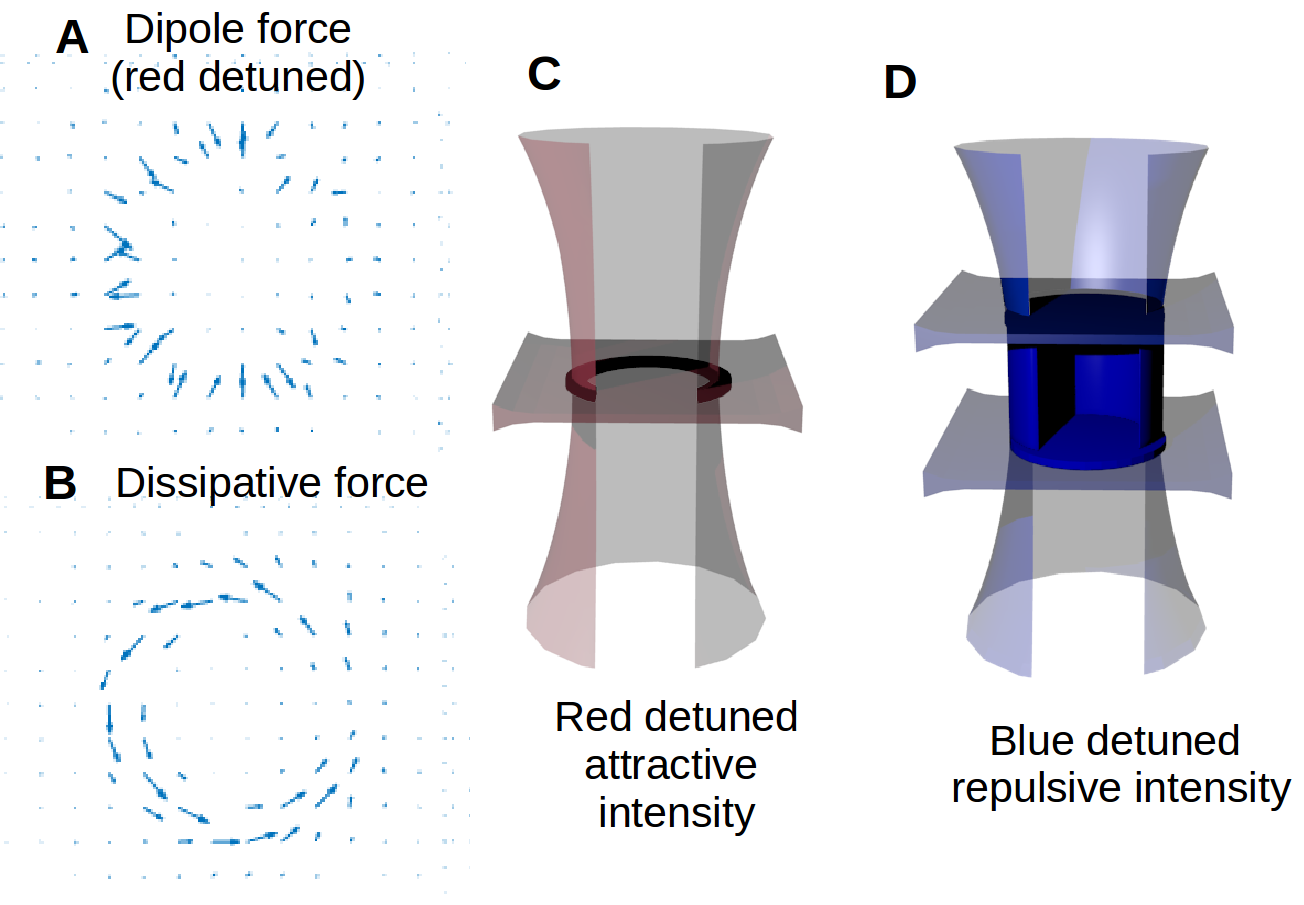}
\caption{{  A) The spatial dependence of the dipole force (arbitrary scale) which would be experienced by an atom inside red detuned C shaped illumination. B)As for A but for the dissipative force.  C) If a red detuned laser is used during atom trapping a C shaped intensity could form an attractive C shaped track for the particle to move along.D) During atom trapping with laser illumination a C shaped beam could form a blue detuned "box" formed by the repulsive intensity barriers.(Gouy rotations are not shown for clarity)}}\label{app}
\end{figure}

Another important area which could benefit from C-shaped beams is in dynamically controllable optical trapping of microparticles and the guiding of atoms in atomtronic device applications.  The phase structure mask for an optical beam can be easily produced, for example, using a spatial light modulator (SLM) technology \cite{Padgett2011a}. The SLM would allow real time variations of the opening angle and/or the size at a speed which would only be limited by the refreshing rate of the SLM. The ability to control the opening angle of an optical trap intensity allows applications such as dynamical adaptive particle sorting. Figure \ref{app} A and B show the force distribution which would be experienced by microparticle.  The effects at play involve trapping along the ring from the dipole potential and rotation around the ring from the optical torque associated with the OAM.

Adjustable C-shaped optical beams can also have interesting applications in the emergent field of atomtronics - the physics and applications of guided current due to gross motion of atoms. For example, if we place a light sheet beam to intersect the C-shaped beam at a right angle (see figure \ref{app}C), the atoms are then trapped in the high intensity regions if the frequencies of both laser beams are red-detuned from the atomic resonance\cite{Chu1986}.  A Bose-Einstein condensate trapped in such a ring circuit with a variable gap is an atomic equivalent quantum interference device\cite{Eckel2014}.  Such a circuitry allows an investigation into tunnelling through the gap barrier, in a manner similar to a Josephson junction\cite{Josephson1962}. Previous experiments have used an additional laser spot to introduce a potential barrier for a Bose-Einstein condensate, however a C-shaped intensity would remove the need for this extra beam thus simplifying the experimental set up. \cite{Ramanathan2011}.

Alternatively, a C-shaped beam intersecting two parallel sheet beams at right angles can act as a cavity (see figure \ref{app}D), if all the laser beams are blue detuned\cite{Davidson1995}.  The SLM can be programmed to control the opening and closing of the cavity to release the trapped atoms in a controlled manner.  

In summary, we have shown how an analytical phase mask function can be used to produce a C-shaped beam embedded with a vortex structure at its centre.  Such a shaped quantum wave has an adjustable opening angle due to the high density of vortex-anti-vortex pairs near the phase discontinuity.  We have demonstrated that the C-shape generated in this way is more robust against beam propagation than the C-shaped beam produced by the conventional Iterative Fourier Transform Algorithm and we have pointed out that this robustness stems from the presence of the topological structure at the core.  A possible application of a C-shaped electron vortex beam would be as a virtual 'stamping' mask in order to reproducibly create nano-structures for applications in metamaterials and plasmonics.  An optical C-shaped beam could, however,  find applications in optical trapping and atomtronics when used to control Bose-Einstein condensates.
\begin{acknowledgements}
This work was carried out with funding from the EPSRC under the grant EP/J022098/1. We would like to thank Dr Michael Ward at the University of Leeds for help with the fabrication of the mask.
\end{acknowledgements}

\bibliographystyle{apsrev4-1}
\bibliography{library,addition_JY}
\newpage
\newpage

\end{document}